\documentclass[prl, twocolumn, floatfix, amsmath, superscriptaddress]{revtex4}
\usepackage{graphicx}
\usepackage{amsmath}
\usepackage{amssymb}
\usepackage{amsthm}
\usepackage{amsfonts}
\usepackage[english]{babel}
\usepackage{url}
\usepackage[colorinlistoftodos]{todonotes}
\providecommand{\bea}{\begin{eqnarray}}
\providecommand{\be}{\begin{equation}}
\providecommand{\eea}{\end{eqnarray}}
\providecommand{\ee}{\end{equation}}
\begin{document}
\title{Optical transmission matrix as a probe of the photonic interaction strength
}

\author{Duygu Akbulut}
\thanks{Present address: ASML, Flight Forum 1900, 5657 EZ Eindhoven, The Netherlands}
\affiliation{Complex Photonic Systems (COPS), MESA+ Institute for Nanotechnology, University of Twente, P.O. Box 217, 7500 AE Enschede, The Netherlands}

\author{Tom Strudley}
\thanks{Present address: Fianium Ltd., Ensign Way, Southampton, UK}
\affiliation{Faculty of Physical Sciences and Engineering, University of Southampton, Highfield, Southampton SO17 1BJ, UK}

\author{Jacopo Bertolotti}
\thanks{Present address: Physics and Astronomy Department, University of Exeter, Stocker Road, Exeter EX4 4QL, UK}
\affiliation{Complex Photonic Systems (COPS), MESA+ Institute for Nanotechnology, University of Twente, P.O. Box 217, 7500 AE Enschede, The Netherlands}

\author{Erik P.A.M. Bakkers}
\affiliation{Department of Applied Physics, TU Eindhoven, Den Dolech 2, 5612 AZ Eindhoven, The Netherlands}
\affiliation{Kavli Institute of Nanoscience, Delft University of Technology, 2600 GA Delft, The Netherlands}

\author{Ad Lagendijk}
\affiliation{Complex Photonic Systems (COPS), MESA+ Institute for Nanotechnology, University of Twente, P.O. Box 217, 7500 AE Enschede, The Netherlands}

\author{Otto L. Muskens}
\affiliation{Faculty of Physical Sciences and Engineering, University of Southampton, Highfield, Southampton SO17 1BJ, UK}

\author{Willem L. Vos}
\affiliation{Complex Photonic Systems (COPS), MESA+ Institute for Nanotechnology, University of Twente, P.O. Box 217, 7500 AE Enschede, The Netherlands}

\author{Allard P. Mosk}
\email{a.p.mosk@utwente.nl}
\affiliation{Complex Photonic Systems (COPS), MESA+ Institute for Nanotechnology, University of Twente, P.O. Box 217, 7500 AE Enschede, The Netherlands}

\date{\today}

\begin{abstract}
We demonstrate that optical transmission matrices (TM) of disordered complex media provide a powerful tool to extract the photonic interaction strength, independent of surface effects. We measure TM of strongly scattering GaP nanowires and plot the singular value density of the measured matrices and a random matrix model. By varying the free parameters of the model, the transport mean free path and effective refractive index, we retrieve the photonic interaction strength. From numerical simulations we conclude that TM statistics is hardly sensitive to surface effects, in contrast to enhanced backscattering or total transmission based methods. 
\end{abstract}
\maketitle
% ----------------------------------------------------------------

Scattering of waves in complex media is a phenomenon of basic scientific interest and of great importance for applications in mesoscopic electron transport, imaging, photovoltaics, lighting, and optical communications~\cite{Ishimaru1978b,Akkermans2007,McCall2009,Wiersma2013}.
In three dimensional (3D) media the interaction strength $S$ is a key parameter \cite{Vos2015} which quantifies how strongly the medium influences the propagation of light waves. In disordered media $S$ quantifies how close a sample is to the Anderson localization transition~\cite{Anderson1958, Lagendijk2009}. 
In the diffusive regime the interaction strength is given by  $S = 1 /k \ell$, with $\ell$ the transport mean free path and $k$ the wave vector inside the medium. At $S \approx 1$ the transition to Anderson localization of ultrasound in 3D  has been observed \cite{Hu2008} and tantalizing indications of a localization transition of light in 3D have emerged \cite{Wiersma1997,Storzer2006}. 
In order to quantitatively study universal properties of scattering media in the diffusive regime and in the transition regime it is of critical importance to have reliable measurements of $S$ in the approach to the transition.

Existing methods to determine $S$ include the measuring of the enhanced backscatter cone~\cite{Albada1985_PRL, Etienne_Wolf1985aa} and of the total transmittance as a function of thickness \cite{Garcia1992}. 
A major limitation to these methods is that they are very sensitive to the inevitable interfaces between the scattering medium and the surroundings with different refractive indices.
In disordered media the interface layer often differs from the bulk, {e.g.}, due to intrinsic sample growth inhomogeneities, exclusion effects, or processing steps.
Hence, the interface contribution may become unpredictable and show large sample to sample variability, which greatly compromises the determination of $S$.

\begin{figure}
\centering
\includegraphics[width=0.85\columnwidth]{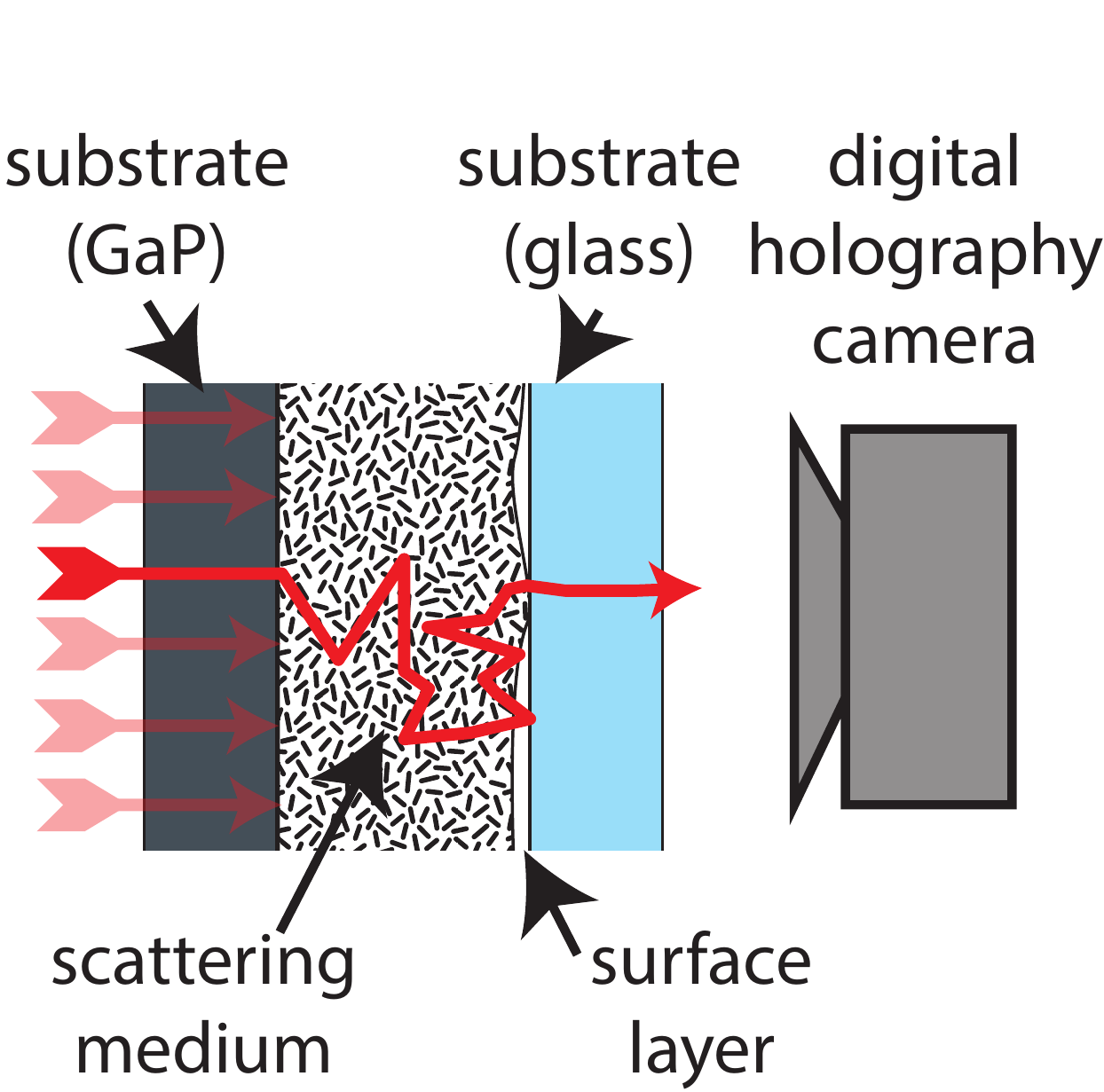}
\caption{(color) 
Cartoon of the transmission matrix measurement. One by one, many non-overlapping incident waves are directed onto a strongly scattering medium. Inside the medium the light diffuses, with contributions from bulk scattering and reflections from the - sometimes ill-defined- surfaces and substrates. The transmitted light is captured by a holographic camera that measures its amplitude and phase.
}
\label{fig:matrix}
\end{figure}

Recently, it has been proposed that the properties of a scattering sample can be sensitively probed through the statistical properties of the transmission matrix (TM)~\cite{Popoff2010aa, Putten2010}. 
The transmission matrix contains the amplitude transmission coefficients between a large number of incident and transmitted modes~\cite{Beenakker1997, Popoff2010aa, Putten2010}, as visualized in Fig.~\ref{fig:matrix}. 
Intensive theoretical studies have been performed on the statistical properties of TMs of disordered waveguides~\cite{Beenakker1997}, which are especially sensitive to the disorder inside the sample.
An important tool in the analysis is the histogram of singular values of the TM. 
In calculations pioneered by Dorokhov, Mello, Pereyra and Kumar (DMPK) \cite{Dorokhov1984, Mello1988a, Imry1986}, this histogram has been found to have a remarkable bimodal shape, containing a high density of singular values that are exponentially small (``closed channels'') as well as some singular values near unity, corresponding to open channels with almost perfect transmission \cite{Pendry1990, Pendry2008phy}. 
Numerical work has confirmed and extended these theoretical results \cite{Muttalib1987,Nazarov1994}. 
For microwave and ultrasound waves TM measurements have confirmed the essential picture of DMPK theory ~\cite{Shi2012, Davy2013OE, Shi2014PNAS, Pena2014NatComm, Gerardin2014_prl}. 
Thus, TM measurements offer a sensitive way to probe the inside of strongly scattering samples. So far, it has not been possible to use this powerful method to obtain quantitative results due to uncertainties about the surface effects.

%%%%%%%%%%%%%%
%%%%%%%%%%%%

In this Letter we report the first use of optical transmission matrix measurements to characterize the scattering medium itself, by probing the scattering strength in the bulk of the medium. 
The normalized singular value histogram of the transmission matrix of a sample (from hereon referred to as \textit{the histogram}) is shown to be a sensitive probe of the bulk scattering strength. 
Through numerical simulations we show that the histogram is insensitive to reflections at the exit interface of the sample, in contrast to other methods. 
From measured transmission matrices of strongly scattering GaP nanowire mats, we obtain a value of $S$ that agrees well with traditional methods.

The results of an optical transport experiment are strongly influenced by the presence of the  interfaces, notably because the refractive index of the scattering medium is usually greater than that of the surrounding medium~\cite{Lagendijk1989, Zhu1991}. 
Light can only exit the medium if the internal angle is smaller than the critical angle, hence both the total transmission and the angular distribution of the transmitted light are changed.
Previous studies of the effect of reflections on the TM employed a tunnel barrier model in a waveguide geometry~\cite{Nazarov1994,Cheng2012_prb}, which does not take into account this angular redistribution of the light. 
Hence the results cannot be directly applied to the optical case.
We have studied the effect of an air layer on the TM, including angular redistribution, in finite-difference time domain calculations (FDTD). 
We calculate the TM of a disordered complex medium with effective index $n_{\rm{eff}} = 1.8$, transport mean free path $\ell=0.6 \,\mu\rm m$ and thickness $L=4 \, \mu$m, with an air layer between the medium and substrate at the transmission side. As a reference, we calculate the TM of the same medium in an index-matched environment \cite{indexmatching2}. 
Remarkably, the width of the histogram is independent of the thickness of the air layer at the exit surface, revealing that it is not sensitive to the interface. 
To meaningfully compare the different methods  we define the apparent photonic interaction strength $S^{\rm app}_i$, which is the result of a measurement of the photonic strength obtained with a certain method ($i$ = TM for TM statistics, EBS for enhanced backscattering, TT for total transmission) that has \emph{not} been corrected for the presence of the air layer \cite{supplement}.
For an ideal bulk-sensitive probe $S^{\rm app}/S$ should be  close to unity  for all interface conditions.

%%%%%%%%%%%%%%
\begin{figure}
\centering
\includegraphics[width=1.0\columnwidth]{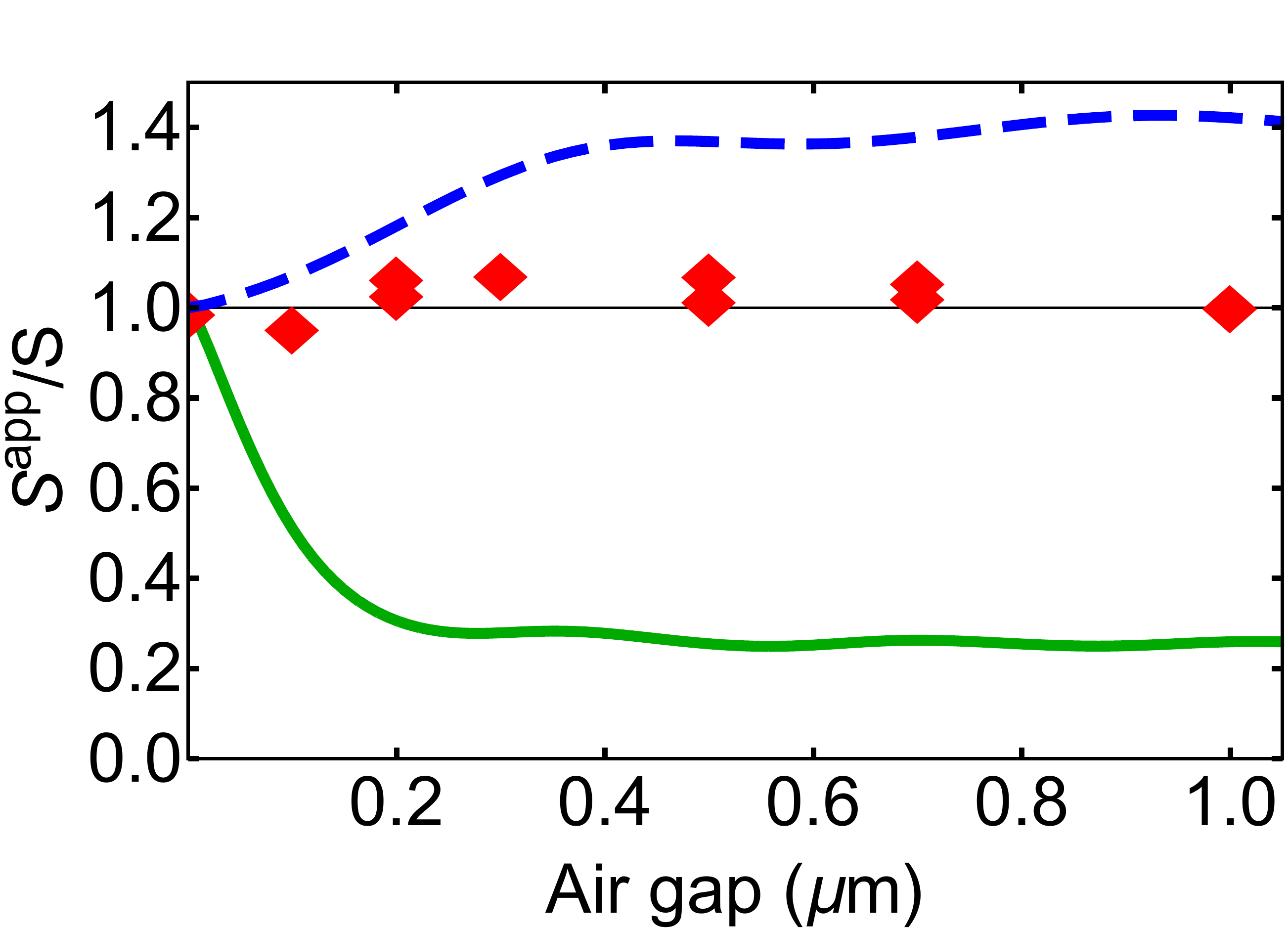}
\caption{(color) Calculated apparent photonic interaction strength for three methods applied to a sample with $L=10\ell$. 
The effective index of the sample is 1.8. 
An air layer ($n=1$) is present between the exit surface of the sample and the index matching medium. 
Red diamonds: $S^{\rm app}_{\rm TM}$ for TM statistics, as calculated by 3D FDTD. 
Blue dashed line: $S^{\rm app}_{\rm TT}$, as calculated for total transmission method. 
Green dotted line: $S^{\rm app}_{\rm EBS}$, as calculated for EBS cone width method~\cite{supplement}.}
\label{fig:threeproxies}
\end{figure}
%%%%%%%%%%%%

In Fig.~\ref{fig:threeproxies} we show the apparent photonic interaction strength for the TM statistics ($S^{\rm app}_{\rm TM}$), as obtained from 3D FDTD calculations, in addition to the corresponding value calculated for backscatter cone and total transmission methods for the same sample geometry. All values are normalized by the true value of $S$.
We see that $S^{\rm app}_{\rm TM}$ is close to $S$ for any air layer thickness.
In contrast, the apparent photonic strengths that result from total transmission ($S^{\rm app}_{\rm TT}$) and from the enhanced backscatter cone width ($S^{\rm app}_{\rm EBS}$) strongly deviate from the true $S$ for air layers thicker than 100 nm.
This deviation indicates that when one uses total transmission or enhanced backscatter cone data to probe the photonic interaction strength, a major correction is required if an air layer or other reflecting surface is present~\cite{supplement}. 
Since the apparent photonic interaction strength for those methods is a steep function of the air layer thickness, a precise knowledge of the exact surface condition is essential to correct for errors in the range of $50 \%$ or even beyond. 
Such precise knowledge is difficult to gain in practice. 
In contrast, in the case of TM statistics, no correction is needed since the apparent $S$ is very close to the true value \cite{FootnoteExitSurface}.
This numerical result demonstrates the important principle that TM statistics is a robust probe of the bulk properties of the complex medium, that is insensitive  even to drastic surface effects such as an air layer.

Disordered semiconductor nanowire mats are extremely strongly scattering samples~\cite{Muskens2009, Strudley2013}. 
Nanowires of GaP, which is a semiconductor material combining transparency in the visible with a very high refractive index of 3.32 at $\lambda$ =632.8 nm~\cite{Aspnes1983aa}, were grown using metal-organic vapor epitaxy on a GaP (100) substrate~\cite{Muskens2009}
to a length of up to 6.4~$\mu$m.
To obtain a maximally disordered arrangement, the nanowires were crushed by applying pressure with a glass slide.  
In samples similar to the ones studied here, a transport mean free path as low as $\ell=0.3~\mu$m at $\lambda=632.8$~nm was observed~\cite{Strudley2013}. 
The effective refractive index  of the nanowire mat is $n_{\rm eff}=$ 1.5 to 2.3, estimated using Bruggeman's  formula~\cite{BohrenandHuffman,supplement}. 
The glass slide was left pressed onto the nanowire mat to allow imaging with an oil immersion objective during the transmission matrix measurements. However, in some samples a sub-$\mu$m air layer of inhomogeneous thickness developed between the nanowires and the glass (see cartoon in Fig.\ref{fig:matrix}). When using TM statistics, even the strong internal reflections caused by such an air layer do not impede  accurate measurements of the bulk scattering strength.

The apparatus used to measure TMs is described in detail in supplementary information~\cite{supplement}. A cartoon is given in Fig.~\ref{fig:matrix}.
Briefly, a spatial light modulator is used to scan the focused spot of a laser (wavelength 633 nm) over the surface of the nanowire layer. 
For each position of the focussed spot the transmitted light field is imaged using off-axis holography~\cite{Leith1962, Takeda1982}. 
Thanks to the use of high-numerical-aperture microscope objectives and by combining measurements of two polarization channels on the incident side, we address 5\% of the incident modes and capture 10\% of the transmitted modes on the (12.8 x 12.8 $\mu$m$^2$) effective area $A$ of the sample.
To measure a large part of the TM, we scanned the incident spot in a checkerboard pattern for each incident polarization. 
The spacing between nearest neighboring spots was 673$\pm$25~nm, which is about one wavelength. 
The incident fields at the front surface of the nanowire layer are measured by repeating the whole measurement procedure with a non-scattering blank sample, which is a GaP slab glued to a glass slide.

\begin{figure}
\centering
\includegraphics[width=1.0\columnwidth]{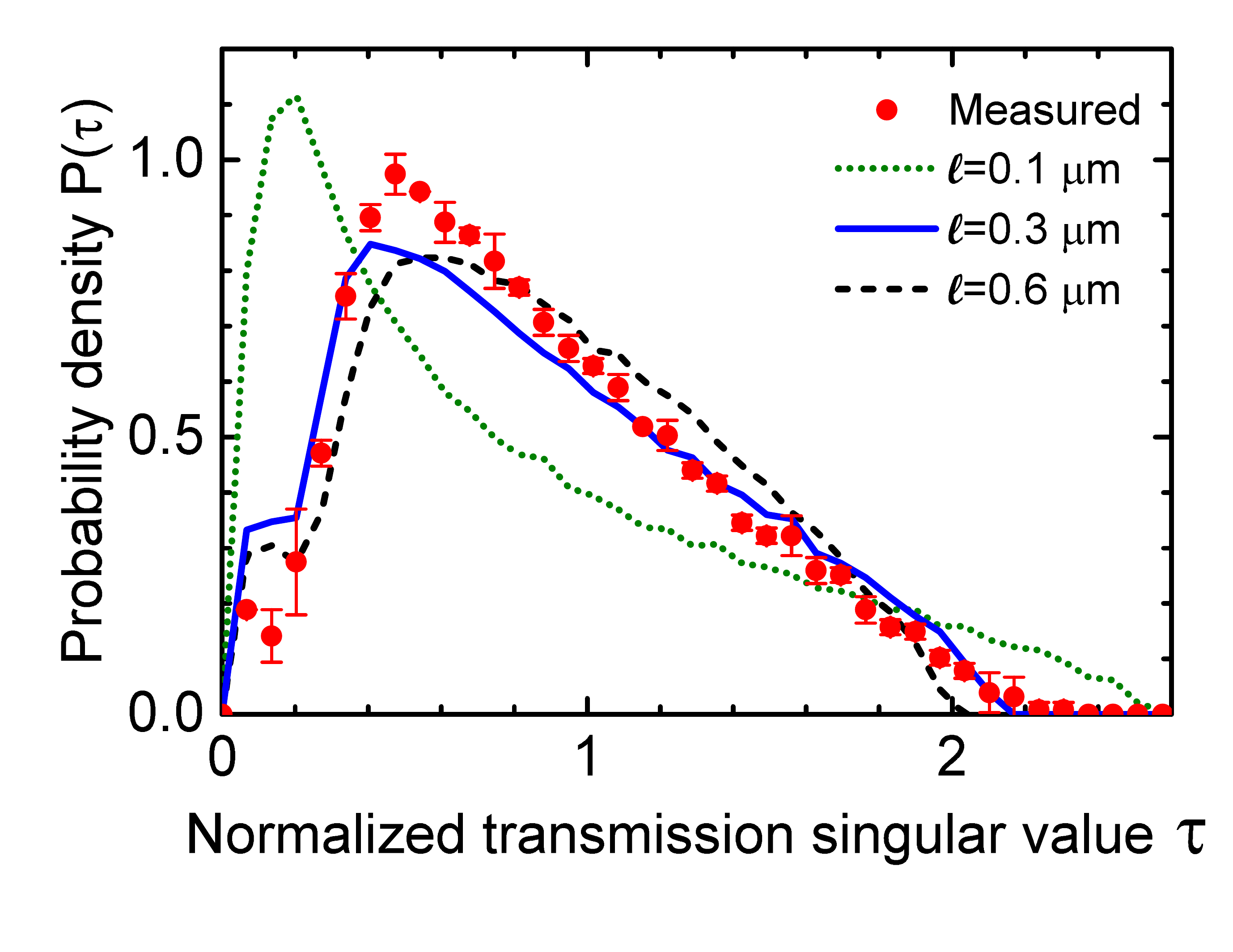}
\caption{(color) Normalized singular value histograms obtained from the experiment (red circles: mean value of 3 experiments; error bars: standard deviation), model with {\it a priori} estimated parameters $\ell=0.3~\mu$m and $n_{\mathrm{eff}}=2.25$ (blue curve), model with larger $\ell=0.6~\mu$m (black curve) and model with smaller $\ell=0.1~\mu$m  (green curve), all with same $n_{\mathrm{eff}}$. 
All model histograms are mean of 20 different histograms generated with independent random matrices.
}
\label{fig:svd_compare_model_exp_NAdetcropped_FOV_13pt9um}
\end{figure}

In Fig.~\ref{fig:svd_compare_model_exp_NAdetcropped_FOV_13pt9um} we show the measured histogram of a disordered GaP nanowire mat.  The singular values $\{\tau\}$ are normalized so that $\langle\tau^2\rangle=1$, \emph{i.e.}, the mean square of the singular values is normalized to unity. 
The histogram has a peak at $\tau=$0.49$^{+0.06}_{-0.05}$ and a slightly concave tail that extends up to $\tau=2.3$.
This histogram is the basis of our quantitative analysis.

To quantitatively compare the measured singular value histogram to theory, we map the waveguide-based DMPK theory to our slab-type samples and take into account the transmission through the optics and the substrate and the fact that in any experiment the TM is {filtered}, \emph{i.e.,} only a finite field of view and only part of the solid angle can be sampled. This filtering strongly affects the shape of the histogram \cite{Olver2012,Goetschy2013}.
We model the internal TM of the sample as a large (8000 $\times$ 8000) matrix with a DMPK singular value density and an average internal transmission of
$
\langle T\rangle=\frac{z_i +z_e}{L+2z_e},
$
where $z_i \approx {\ell}$ is the effective injection depth, and $z_e \approx 2 {\ell}/3$ is the extrapolation length~\cite{GomezRivas1999, Akkermans1986aa, supplement}.
The effective filtering ratio follows from approximating the sample as a waveguide with a cross-section area $A$, with the width taken as the average of the width of the probed area on the incident surface and the FWHM width the diffuse transmitted spot.
On the input side, the filtering ratio is the ratio of the number of probed modes to $N_{\rm wg}=\frac{2\pi A n^{2}_{\mathrm{eff}}}{\lambda^{2}}$, with $\lambda$ being the free space wavelength~\cite{Gloge1971_ao}. 
On the output side the filtering is due to the detection N\!A, therefore the filtering ratio is ${\rm N\!A}^2/2n_{\rm eff}^2$. The factor 2 in the denominator is due to detection being made for a single polarization. As the filtering is asymmetric, the TM is rectangular.
Finally, we take into account the propagation through the optics and substrate by multiplying the model TM by the measured transmission matrix $T_0$ of a non-scattering reference sample~\cite{supplement}. 
By this procedure any mode overlap introduced by our optics and field generation are included in the model. 
Therefore, the histograms from the experiment can be compared directly and quantitatively to those of the model.

%%%%%%%%%%%%%%
%%%%%%%%%%%%

The histogram obtained from the model for a realistic estimate of the sample parameters, $n_{\mathrm{eff}}=2.25$ and $\ell=0.3~\mu$m is shown in Fig.~\ref{fig:svd_compare_model_exp_NAdetcropped_FOV_13pt9um}, along with the histograms obtained for an unrealistically high $\ell=0.6~\mu$m and for an unrealistically low $\ell=0.1~\mu$m, while retaining the same estimate for $n_{\mathrm{eff}}$. 
The model and the experimental histograms are in good agreement for $\ell=0.3~\mu$m. Both curves are asymmetric in shape with a sharp rise at low singular values to reach a peak at $\tau \approx 0.4$.
After the peak, both histograms decrease in a slightly concave manner, with the experimental histogram having a higher slope, both reaching 0 counts at $\tau\approx 2.3$. 
The model histogram with the longer mean free path shows an obviously more convex shape than the experimental data, and the model histogram with the shorter mean free path is more pronouncedly concave, indicating that the scattering strength can be read directly from the shape of the histogram.

%%%%%%%%%%%%
\begin{figure}
\centering
\includegraphics[width=1.0\columnwidth]{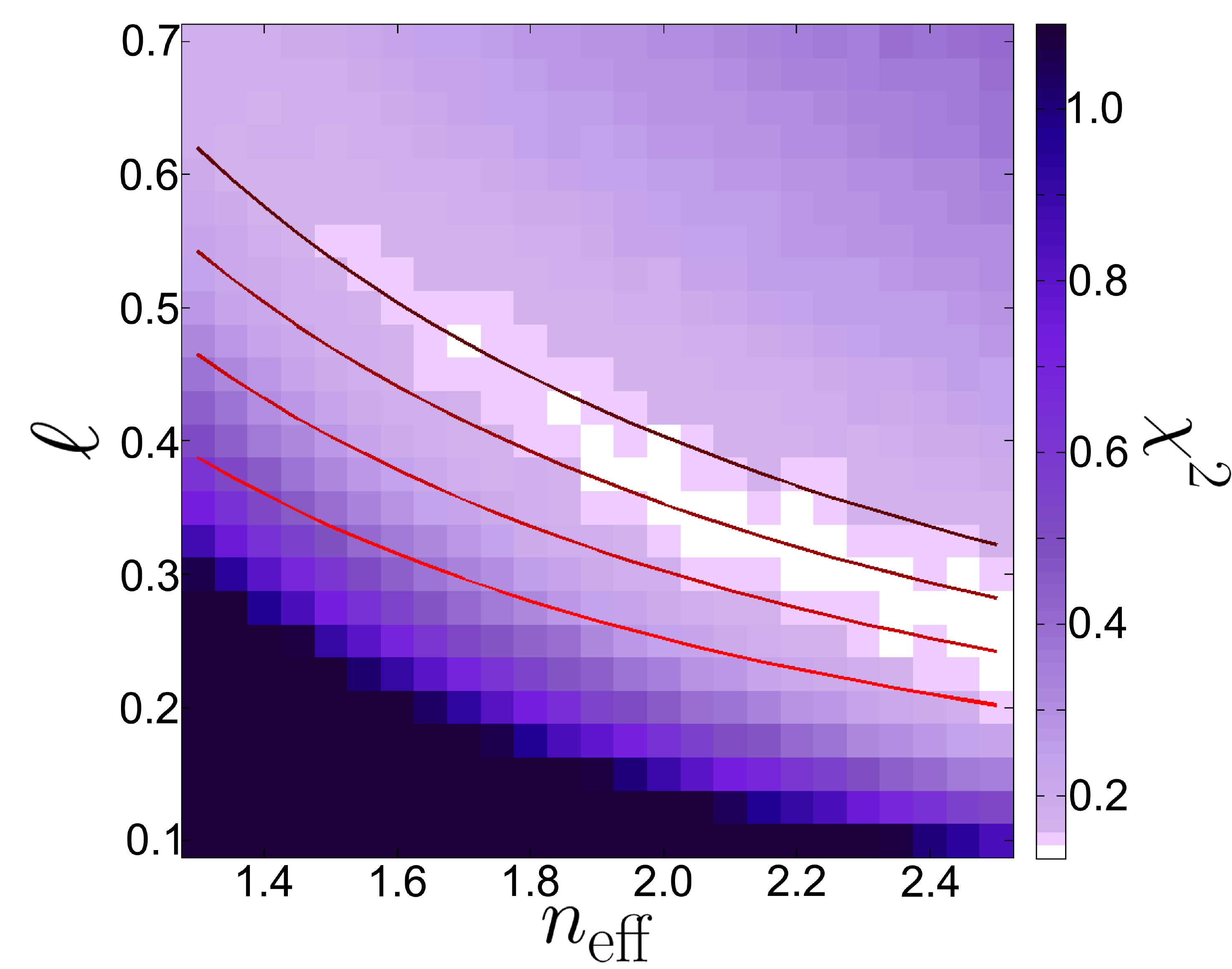}
\caption{ (color)
Map of $\chi^{2}$, the distance metric between the numerical model and the experimental data, as a function of the parameters $\ell$ and $n_{\mathrm{eff}}$. The color scale saturates at $\chi^{2}=1.1$. The white region corresponds to the best agreement between experiment and the model. Red curves: $k\ell=$5 to $k\ell=$8 (incremented by 1 for each curve from light red to dark red).
}
\label{fig:GaPparametermaps}
\end{figure}
%%%%%%%%%%%%

We now describe the procedure we used to retrieve the photonic strength. 
The effective area $A$ and the length $L$ of the sample as well as the number of probed modes on incident and outgoing sides are fixed parameters of our model, while the mean free path $\ell$ and the effective index $n_{\mathrm{eff}}$ are {\it a priori} adjustable parameters.
In Fig.~\ref{fig:GaPparametermaps} we show the distance metric $\chi^{2}$ between model and experimental data for a rectangular domain encompassing the possible range of the transport mean free path $\ell$ and the effective index $n_{\mathrm{eff}}$. 
The region of minimum $\chi^{2}$ is a diagonal valley running from high transmission and low effective index to low transmission and high effective index. 
This valley approximately tracks the curves of constant $S$. 
Hence, while the comparison between the model histograms and the experimental histograms gives little independent information on ${\ell}$ and $n_{\mathrm{eff}}$, we find that the photonic interaction strength can be accurately determined as $S$$=$$0.14$, or equivalently, $k\ell$$=$$7$, with a 20\% error estimate.
This error margin is determined by considering the minimum and maximum of $S$ in the region where $\chi^{2}-\chi_{\rm min}^{2}$$<$$3\sigma$, where $\chi_{\rm min}^{2}$ is the global minimum and $\sigma$ is the standard deviation of $\chi_{\rm min}^{2}$ as obtained from the comparison between the average experimental histogram and each model histogram.
This procedure estimates the statistical error due to the parameter estimation procedure and the slight deviation of the valley of best fit from the constant $S$ curves. Furthermore the uncertainty in the thickness of the sample, which is on the order of 8\%, is included in the 20\% error estimate.
The value of $S$ obtained here is compatible with the measurements reported in Ref.~\cite{Strudley2013} and the $n_{\mathrm{eff}}$ values estimated from the filling fraction.
The level of uncertainty reached here is good compared to other methods of measuring the photonic interaction strength, such as enhanced backscattering or total transmission measurements, considering that no {\it a priori} assumption about $n_{\mathrm{eff}}$ is made and that the method is not sensitive to surface effects.

In summary, we have demonstrated that the measurement of the transmission matrix is a powerful method to characterize the properties of a scattering material, and that this approach is surprisingly robust to common artifacts related to internal reflections by surface layers. 
In particular, we have shown that the transmission matrix measurements can be modeled with wave transport theory to reliably yield the photonic interaction strength as the only relevant free parameter. 
The method is therefore very well suited to investigate mesoscopic samples with rough surfaces such as photonic glasses \cite{Garcia2007}, powders
and disordered photonic band gap crystals, as well as 3D ultrasound media \cite{Hu2008}. A precise characterization of the bulk scattering strength is a prerequisite to a quantitative understanding of the Anderson transition in such media.

We thank Pepijn Pinkse, Klaus Boller, Arthur Goetschy and Douglas Stone for insightful discussions, and Cornelis Harteveld for technical assistance.
This work is part of the research program of the ``Stichting voor Fundamenteel Onderzoek der Materie (FOM)'', which is financially supported by the ``Nederlandse Organisatie voor Wetenschappelijk Onderzoek (NWO)''.
We acknowledge support from ERC grant 27948, NWO-Vici, STW, the Royal Society, and EPSRC through fellowship EP/J016918/1.

\bibliographystyle{apsrev}
\bibliography{GapManuscript}

\begin{thebibliography}{46}
\bibitem[{ind()}]{indexmatching2}
\bibinfo{note}{The index of the matching environment in the numerical
  calculations is 2.0, slightly higher than $n_{\rm eff}$, which leads to
  negligible internal reflection,}.
\end{thebibliography}

 ----------------------------------------------------------------
% ----------------------------------------------------------------

\end{document}